\documentclass[aps,showpacs,preprint,superscriptaddress]{revtex4}
\usepackage{graphicx}
\usepackage{subfigure}
\usepackage{float}
\usepackage{amsmath}
\usepackage{amsfonts}
\usepackage{color}
\usepackage{bm}
\usepackage{bbm}
\usepackage{txfonts}
\usepackage{array}
\usepackage{amssymb}

\begin{document}

\title{Effects of finite spatial extent on Schwinger pair production}
\author{Mamutjan Ababekri}
\affiliation{College of Mathematics and System Sciences, Xinjiang University, Urumqi 830046, China}
\author{Bai-Song Xie \footnote{bsxie@bnu.edu.cn}}
\affiliation{Key Laboratory of Beam Technology of the Ministry of Education, and College of Nuclear Science and Technology, Beijing Normal University, Beijing 100875, China}
\affiliation{Beijing Radiation Center, Beijing 100875, China}
\author{Jun Zhang \footnote{zhj@xju.edu.cn}}
\affiliation{School of Physics Science and Technology, Xinjiang University, Urumqi 830046, China}

\date{\today}

\begin{abstract}
Electron-positron pair production from vacuum in external electric fields with space and time dependencies is studied numerically using real time Dirac-Heisenberg-Wigner formalism.
The influence of spatial focusing scale of the electric field on momentum distribution and the total yield of the particles is investigated by considering space-time dependent electric fields in 1+1 dimensions with various temporal configurations.
With the decrease of spatial extent of the external field, signatures of the temporal field are weaken in the momentum spectrum.
Moreover, in the extremely small spatial extent, novel features emerge due to the combined effects of both temporal and spatial variations.
We also find that for dynamically assisted particle production, while the total particle yield drops significantly in small spatial extents, the assistance mechanism tends to increase in these highly inhomogeneous regimes, where the slow and fast pulses are affected differently by the overall spatial inhomogeneity.
\end{abstract}
\maketitle

\section{Introduction}

Schwinger pair production is a nonperturbative phenomenon in quantum electrodynamics(QED) in which electrons and positrons are created from vacuum under intense electromagnetic fields\cite{Sauter:1931zz,Heisenberg:1935qt,Schwinger:1951nm}.
This highly nontrivial prediction of QED is exponentially suppressed due to its tunneling nature and its detection has remained as a challenge for many decades.
However, advances in high intensity laser technology in recent years and the upcoming experiments \cite{Ringwald:2001ib,Heinzl:2008an,Marklund:2008gj,Pike:2014wha} have brought the hope of observing pair production in the laboratory and spurred the interest in studying pair production under intense fields.
Understanding the nature of pair production in the nonperturbative regime would not only deepen our knowledge about the relatively less tested branch of QED \cite{Dunne:2008kc}, but it may also shed light on other nonperturbative phenomena like Unruh effect and Hawking radiation where direct detections are not possible \cite{Schutzhold:2010rq}.

More complex external fields have been investigated and great progress is made since the original works on Schwinger pair production where static electric field was considered, for a recent review see Ref. \cite{Gelis:2015kya}.
In addition to experimental considerations, one has to explore more realistic laser fields for the theoretical implication that physical observable of pair production depends nontrivially on external field parameters \cite{Hebenstreit:2009km,DiPiazza:2009py,Dumlu:2010vv}.
Because of this strong sensitivity of pair creation to external field shapes, which is expected for such nonlinear effect, adding spatial inhomogeneity into external fields is inevitable to provide more reliable predictions in the study of Schwinger pair production.
With the development in the theory and breakthrough in computational techniques, various consequences of spatial dependency of the external fields have been studied in different theoretical approaches in recent years \cite{Gies:2005bz,Dunne:2006ur,Ruf:2009zz,Heinzl:2010vg,Schneider:2014mla,Wollert:2015kra,Wollert:2016fwp,Dumlu:2013pma,Dumlu:2015paa,Aleksandrov:2017mtq,Lv:2018wpn,Hebenstreit:2011wk,Kohlfurst:2017hbd}.
In the frame work of real time Dirac-Heisenberg-Wigner(DHW) formalism\cite{Vasak:1987um,Hebenstreit:2011wk}, particle self-bunching effect is discovered for the standing wave profile with finite spatial extension\cite{Hebenstreit:2011wk}.
Moreover, when finite spatial pulse size is introduced to the multi-photon pair production process, it is shown that the ponderomotive force due to strong spatial focusing causes peak splitting in the momentum spectrum \cite{Kohlfurst:2017hbd}.
These findings imply that the spatial inhomogeneity of the external fields, combined with various temporal field profiles, affects produced particle's momentum spectrum as well as production rate and may result in new phenomena, see also Ref. \cite{Aleksandrov:2017mtq}.

One of the profound features of Schwinger pair production is the interference effect shown to be present in the momentum distribution of created particles in temporal pulses with subcycle structure or multiple pulses of alternating signs \cite{Hebenstreit:2009km,Akkermans:2011yn,Kaminski:2018ywj}.
Such interference effects can be explained semiclassically in terms of interference between pairs of turning points in the complex plane \cite{Dumlu:2010ua,Dumlu:2011rr}.
Another significant result is the dynamically assisted Schwinger mechanism which brought the hope of observing vacuum pair creation below the Schwinger critical field strength with the upcoming experimental setups\cite{Schutzhold:2008pz}.
The strong pulse lowers the threshold for particle creation induced by the weak pulse, thus the combined result is greater than the contribution of the added results of particle creation from individual pulse.
It is worth investigating to what extent various spatial variation scales would change these results where detailed structure of the external pulses are crucial, see Ref. \cite{Heinzl:2010vg} for related investigations in laser induced pair production.

In this paper, by adopting the numerical techniques developed in \cite{Kohlfurst:2015zxi}, we use the real time DHW formalism to investigate how the results of vacuum pair production are affected by the finite spatial extent of the electric field by considering space and time variations for the external field with different temporal structures.
We note that the momentum spectrum signatures of temporal pulses are largely influenced by highly inhomogeneous cases where the spatial extent is small, and particularly around $2 \lambda_c$(the electron Compton wavelength) novel features may emerge due to the interplay between the temporal and spatial structures.
In addition, the overall spatial inhomogeneity in dynamical assistance mechanism affects the fast and slow pulses differently causing the assistance factor to vary for different values of spatial extents.

Our paper is organized as follows.
In Sec.~\ref{method}, we explain the treatment of Schwinger pair production under 1+1 dimensional space-time dependent electric fields:
After the introduction of the external field to be considered in our paper in Sec.~\ref{fields}, we briefly review DHW formalism used in our work and discuss the numerical strategy of solving the partial differential equations in Sec.~\ref{DHWformalism} and Sec.~\ref{strategy}, respectively.
In Sec.~\ref{results}, the numerical results obtained for various filed forms are presented with physical discussions.
Sec.~\ref{result1}, Sec.~\ref{result2} and Sec.~\ref{result3} show the momentum spectrum for single pulse, two same signed pulses, and two opposite signed pulses, respectively.
In Sec.~\ref{result4}, the effect of spatial constraint on assistance mechanism is presented.
The summary and outlook are given in Sec.~\ref{summary}.

Throughout this article, natural units($ \hbar = c = 1 $) are used and the quantities are presented in terms of the electron mass $m$ and the electron Compton wavelength $\lambda_c$.

\section{Pair production in 1+1 dimensions}\label{method}

\subsection{External fields}\label{fields}
In the present work, we study electron positron pair production in 1+1 dimensions by considering the following idealized electric field mode with space and time dependencies:
\begin{equation}\label{FieldMode}
\begin{aligned}
E\left(x,t\right)
&=E_{0} f \left( x \right ) g\left( t \right )\\
&=\epsilon \, E_{cr} \exp \left(-\frac{x^{2}}{2 \lambda^{2}} \right ) g\left(t\right),
\end{aligned}
\end{equation}
where $E_{cr}$ is the critical field strength.
Throughout this paper, the spatial part $f \left( x \right)$ takes the Gaussian shape while we assign various forms for the temporal pulse whose detailed structure will be presented in due sections.
By doing so, we are introducing into the external field a finite spatial extent which is scaled by the value of $\lambda$.
Also note that the field strength varies with $x$ and $t$, and the direction will be along the $x$-axis.
Therefore, we achieve computational advantage and the magnetic field vanishes in this simplistic model for two counter-propagating lasers.
In this simplified field mode Eq.\eqref{FieldMode}, we are investigating how spatial variation affects pair production by calculating the particle momentum distribution and total yield at asymptotic times $t \rightarrow \infty$ for different $\lambda$.

\subsection{DHW formalism}\label{DHWformalism}
To study vacuum pair production in both space and time dependent electric fields given Eq. \eqref{FieldMode}, we will be employing the real time DHW formalism \cite{Vasak:1987um,Hebenstreit:2011wk}, which is shown to be quite useful in the study of pair production \cite{Hebenstreit:2011wk,Hebenstreit:2013qxa,Kohlfurst:2015niu,Blinne:2013via,Li:2015cea,Olugh:2018seh}, especially in the case of spatially inhomogeneous external fields \cite{Hebenstreit:2011wk,Kohlfurst:2015niu,Kohlfurst:2017hbd}.
In the following, we present brief review of the DHW formalism as the framework adopted to our study.

The main ingredients of the DHW formalism are Dirac spinor fields $\psi$,$\bar{\psi}$, and the vector potential $\mathbf{A}\left(x,t\right)$, where the latter is considered to be a classical external field as an apt approximation in the study of Schwinger pair production.
We start by introducing the density operator using the commutator for Dirac spinors,
\begin{equation}\label{DensityOperator}
 \hat {\mathcal C}_{\alpha \beta} \left( r , s \right) = \mathcal U \left(A,r,s
\right) \ \left[ \bar \psi_\beta \left( r - s/2 \right), \psi_\alpha \left( r +
s/2 \right) \right],
\end{equation}
where $r$ denotes the center-of-mass  and $s$ the relative coordinate.
The Wilson line factor is used to ensure gauge invariance:
\begin{equation}
 \mathcal U \left(A,r,s \right) = \exp \left( \mathrm{i} \ e \ s \int_{-1/2}^{1/2} d
\xi \ A \left(r+ \xi s \right)  \right).
\end{equation}
Now the covariant Wigner operator could be defined as a Fourier transform of the density operator Eq.~\eqref{DensityOperator},
\begin{equation}\label{WignerOperator}
 \hat{\mathcal W}_{\alpha \beta} \left( r , p \right) = \frac{1}{2} \int d^4 s \
\mathrm{e}^{\mathrm{i} ps} \  \hat{\mathcal C}_{\alpha \beta} \left( r , s
\right).
\end{equation}
By taking the vacuum expectation value of Eq.~\eqref{WignerOperator}, we obtain the following covariant Wigner function,
\begin{equation}
 \mathbbm{W} \left( r,p \right) = \langle \Phi \vert \hat{\mathcal W} \left( r,p
\right) \vert \Phi \rangle.
\end{equation}
Since we are treating the background field as a complex number valued function --- consequence of the Hartree type approximation, the electromagnetic field can be factored out:
\begin{equation}
 \langle \Phi \vert F_{\mu \nu} \ \hat{\mathcal{C}} \vert \Phi \rangle =
F_{\mu \nu} \langle \Phi \vert \hat{\mathcal{C}}
\vert \Phi \rangle \, .
\end{equation}

Considering the fact that the Wigner function is in the Dirac algebra, it could be rewritten in terms of 16 covariant Wigner coefficients:
\begin{equation}
\mathbbm{W} = \frac{1}{4} \left( \mathbbm{1} \mathbbm{S} + \textrm{i} \gamma_5
\mathbbm{P} + \gamma^{\mu} \mathbbm{V}_{\mu} + \gamma^{\mu} \gamma_5
\mathbbm{A}_{\mu} + \sigma^{\mu \nu} \mathbbm{T}_{\mu \nu} \right) \,
\label{decomp}
\end{equation}
where the transformation properties of each component are implied by their notations.
Because we are dealing with particle pair creation and we want to describe it as an initial value problem, we switch to the equal-time formalism which can be achieved by taking the energy average of the Wigner function,
\begin{equation}
 \mathbbm{w} \left( \mathbf{x}, \mathbf{p}, t \right) = \int \frac{d p_0}{2 \pi}
\ \mathbbm{W} \left( r,p \right)
\end{equation}
where $\mathbf{x}$ and $\mathbf{p}$ represent the position and kinetic momentum of the particles.

The derivation of the equations of motion using the Dirac equation yields the 3+1 dimensional results which could be further reduced to the presently considered 1+1 dimensional ones by considering $\mathbf{A}\left(\mathbf{x},t\right )=A\left(x,t\right )\mathbf{e}_{x}$:
\begin{align}
 &D_t \mathbbm{s} - 2 p_x \mathbbm{p} = 0 , \label{pde:1}\\
 &D_t \mathbbm{v}_{0} + \partial _{x} \mathbbm{v}_{1} = 0 , \label{pde:2}\\
 &D_t \mathbbm{v}_{1} + \partial _{x} \mathbbm{v}_{0} = -2 m \mathbbm{p} , \label{pde:3}\\
 &D_t \mathbbm{p} + 2 p_x \mathbbm{s} = 2 m \mathbbm{v}_{1} , \label{pde:4}
\end{align}
with the pseudo-differential operator
\begin{equation}\label{pseudoDiff}
 D_t = \partial_{t} + e \int_{-1/2}^{1/2} d \xi \,\,\, E_{x} \left( x + i \xi \partial_{p_{x}} \, , t \right) \partial_{p_{x}} .
\end{equation}
And the vacuum initial conditions are \cite{Kohlfurst:2015zxi}:
\begin{equation}\label{vacuum-initial}
{\mathbbm s}_{vac} = - \frac{2m}{\omega} \, ,
\quad  {\mathbbm v}_{1 \, vac} = - \frac{2{ p_x} }{\omega} \,  ,
\end{equation}
where $\omega$ is the energy of a particle defined as $\omega=\sqrt{p_{x}^{2}+m^2}$.
The vacuum initial states become zero if we redefine the Wigner components
\begin{equation}
\mathbbm{w}_{k}^{v} \left( x , p_{x} , t \right) = \mathbbm{w}_{k} \left( x , p_{x} , t \right) - \mathbbm{w}_{vac} \left(p_{x} \right),
\end{equation}
where $\mathbbm{w}_{k}$ is the Wigner component in Eqs. \eqref{pde:1}-\eqref{pde:4} (with the correspondence: $\mathbbm{w}_{0} = \mathbbm{s}$ , $\mathbbm{w}_{1} = \mathbbm{v}_{0}$ , $\mathbbm{w}_{2} = \mathbbm{v}_{1}$ and $\mathbbm{w}_{3} = \mathbbm{p}$), and $\mathbbm{w}_{vac}$ is the corresponding vacuum initial condition \eqref{vacuum-initial}.
The particle number density in the momentum space is defined as follows:
\begin{equation}\label{MS}
n \left( p_{x} , t \right) = \int \frac{ d x}{2 \pi}  \, \frac{\mathbbm{s}^{v} \left( x , p_{x} , t \right) + p_{x} \, \mathbbm{v}_{1}^{v} \left( x , p_{x} , t \right)}{\omega \left( p_{x} \right)}.
\end{equation}
Then the total particle yield is given by:
\begin{equation}\label{totalYield}
N \left( t \right) = \int d p_{x} \, n \left( p_{x} , t \right).
\end{equation}

\subsection{Numerical strategy}\label{strategy}
The challenging part in solving the partial differential equations (PDEs) \eqref{pde:1}-\eqref{pde:4} is the pseudo-differential operator \eqref{pseudoDiff} which, for convenience, can be split into two parts: $D_{t} = \partial_{t} + \Delta$, and the PDEs \eqref{pde:1}-\eqref{pde:4} then take the final form:
\begin{align}
&\partial_{t} \mathbbm{w}_{0}^{v} = - \Delta \mathbbm{w}_{0}^{v} + 2 p_{x} \mathbbm{w}_{3}^{v}, \\
&\partial_{t} \mathbbm{w}_{1}^{v} = - \Delta \mathbbm{w}_{1}^{v} - \partial_{x} \mathbbm{w}_{2}^{v}, \\
&\partial_{t} \mathbbm{w}_{2}^{v} = - \Delta \mathbbm{w}_{2}^{v} - \partial_{x} \mathbbm{w}_{1}^{v} - 2 \mathbbm{w}_{3}^{v}, \\
&\partial_{t} \mathbbm{w}_{3}^{v} = - \Delta \mathbbm{w}_{3}^{v} - 2 p_{x} \mathbbm{w}_{0}^{v} + 2 \mathbbm{w}_{2}^{v}.
\end{align}

We can apply the operator $\Delta$ on Wigner functions by employing the Fourier transform and inverse Fourier transform,
\begin{equation}\label{FT-IFT}
f\left ( p_{x}  \right ) = \mathcal{F}^{-1}_{p_{x}} \left [ \mathcal{F}_{p_{x}} \left [ f \left( p_{x} \right ) \right ] \right ] = \mathcal{F}^{-1}_{p_{x}} \left[ \tilde{f}\left(\text{k}_{p_{x}} \right) \right] ,
\end{equation}
with the Taylor expansion of the electric field,
\begin{equation}\label{Taylor}
\mathcal{F}_{p_{x}} \left [ \frac{d^{n}}{d p_{x}^{n}} f \left( p_{x} \right ) \right ] = \left ( i \text{k}_{p_{x}} \right )^{n} \tilde{f}\left(\text{k}_{p_{x}}\right),
\end{equation}
and resumming the expansion to obtain the final form \cite{Kohlfurst:2015zxi},
\begin{equation}\label{WignerFT-IFT}
\Delta \mathbbm{w}_{k}\left( x , p_{x} , t \right )=\mathcal{F}^{-1}_{p_{x}}\left[i e \text{k}_{p_{x}} \int d \xi E \left( x - \xi \text{k}_{p_{x}} , t\right) \mathbbm{w}_{k}\left(x, \text{k}_{p_{x}} ,t \right) \right].
\end{equation}

Note that, since the Fourier transform takes all points of the domain into account, we may obtain high resolution results in whole phase space \cite{Kohlfurst:2017hbd}.
Also, the spatial and momentum directions are equidistantly discretized, transforming the PDEs into ($N_x \times N_{p_{x}}$) systems of ODEs that could be solved readily \cite{Kohlfurst:2017hbd,Kohlfurst:2015zxi}.

\section{Numerical results}\label{results}
In this section, by adopting the numerical method introduced in the previous section, we study the $e^{+}e^{-}$ pair production in 1+1 dimensions with field forms given in Eq. \eqref{FieldMode}.
As mentioned in Sec.~\ref{fields}, we will be changing the spatial focusing size and obtain results for different temporal modes including single temporal pulse, two symmetric and anti-symmetric pulses, and the dynamically assisted pulse.

\subsection{One Sauter pulse}\label{result1}

\begin{figure}[h]\suppressfloats
\includegraphics[scale=0.6]{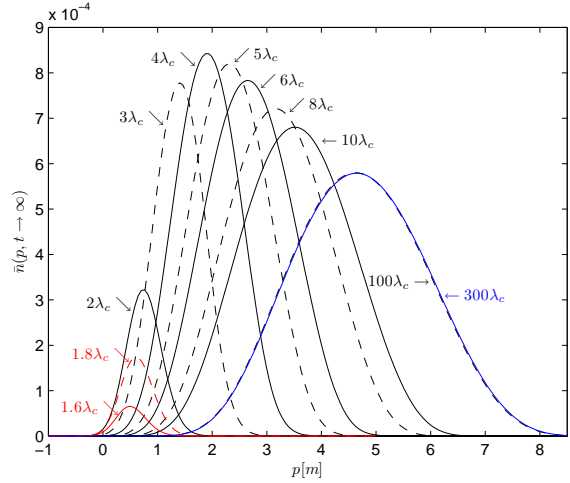}
\caption{Reduced momentum spectrum for different values of spatial extents for the single pulse  \eqref{FieldMode0} with $ E_{0}=0.5 E_{cr}$ and $\tau=10$m$^{-1}$. The black lines in this figure is same as Fig. 2 of Ref. \cite{Hebenstreit:2011wk} except for the additional lines for $\lambda = 300 \lambda_c$(the blue line), $\lambda = 1.8 \lambda_c$ and $\lambda = 1.6 \lambda_c$(two red lines) that are added to assist our analysis.}
\label{self-bunching}
\end{figure}

We start by discussing the effects of spatial inhomogeneity at various spatial extents in the single temporal pulse scenario.
To do so, we choose $g \left( t \right ) = \text{sech}^{2}\left( \frac{t}{\tau} \right) $ in Eq. \eqref{FieldMode}, and the external electric field reads:
\begin{equation}\label{FieldMode0}
E\left(x,t\right) = E_{0} \exp \left(-\frac{x^{2}}{2 \lambda^{2}} \right ) \text{sech}^{2}\left( \frac{t}{\tau} \right).
\end{equation}
We calculate the reduced observable $\bar{n}\left( p , t \right) = n \left( p , t \right) / \lambda$ and $ \bar{N}\left( t \right ) = N \left( t \right ) / \lambda$, defined as the momentum spectrum \eqref{MS} and total yield \eqref{totalYield} divided by pulse length $\lambda$ of the spatial part of the electric field, to extract the nontrivial effects of spatial inhomogeneity for different values of spatial extents\cite{Hebenstreit:2011wk}.
We are assuming the only nonzero momentum is $p_x$ and set $p \equiv p_x$.

To assist our analysis of the results obtained in the following sections, we provide Fig. \ref{self-bunching} in our article with a few more curves compared to the corresponding figure in Ref. \cite{Hebenstreit:2011wk}, where the particle self-bunching effect is first predicted.

As can be seen from Fig. \ref{self-bunching}, the uniform field approximation is already valid for $\lambda = 100 \lambda_c$ such that the electric field can be considered quasihomogeneous and momentum spectrum does not change for larger spatial extents.
At these quasihomogeneous regimes, we may expect the special features of pure temporal pulses are recovered with slight differences.
In the case of the uniform field, particles created at different temporal regions could interfere with each other and could be affected by the temporal filed; however, this may change if we have spatial dependence since there are always some particles could escape from the external field regions with no further interaction with other particles or the external field.
In addition, because of the Gaussian shape of the spatial dependency of the external field, the electric field is not uniform even at the considered largest extension scales.
We can observe such effects in Fig. \ref{Np-Di1Sauter-new} and Fig. \ref{Np-DA}, where the interference and the dynamical assistance features at the quasihomogeneous limit are still less profound in the momentum spectrum compared to the pure temporal counterparts.

Here, we have plotted momentum spectrum curves for $\lambda$ taking values down to $1.6 \lambda_c$, since it is within the estimated minimum spatial extent, for the considered field form, beyond which pair creation terminates \cite{Hebenstreit:2011wk}:
\begin{equation}\label{mini-lambda}
\lambda \ge \frac{E_{cr}}{E_0} \sqrt{\frac{2}{\pi}} \lambda_c \approx 1.6 \lambda_c.
\end{equation}
Note that momentum spectrum curves keep the vanishing tendency for $\lambda \le 2 \lambda_c$.

When multiple temporal pulses are considered, as will be shown in the following sections, we will see different results for various scales of spatial variations from the quasihomogeneous case to the extremely small spatial extent which may introduce novel phenomena in the momentum spectrum.

\subsection{Two symmetric Sauter pulses}\label{result2}

\begin{figure}[h]\suppressfloats
\includegraphics[scale=0.6]{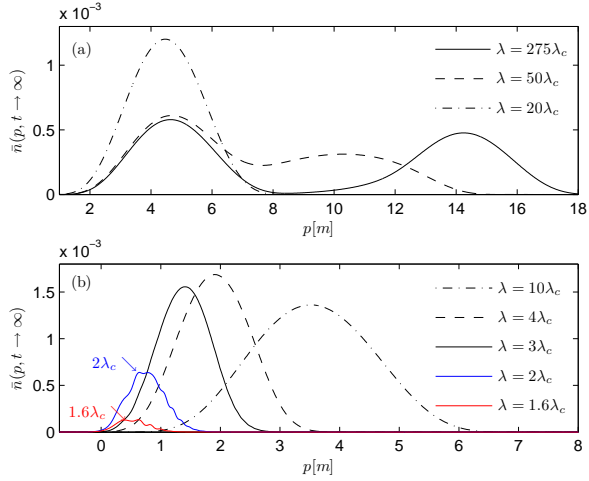}
\caption{Momentum spectrum for different spatial extents for two symmetric pulses \eqref{FieldMode2} with $ E_{0}=0.5  E_{cr}$ , $\tau=10$m$^{-1}$ and $\Delta=3$.}
\label{Np-Double3Sauter}
\end{figure}

\begin{figure}[h]\suppressfloats
\includegraphics[scale=0.6]{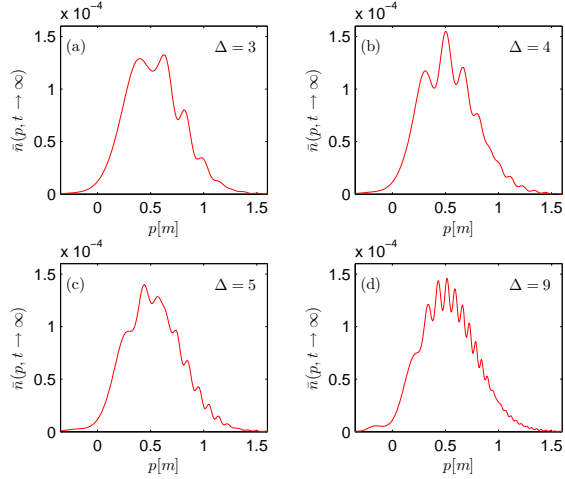}
\caption{Momentum spectrum for two symmetric pulses \eqref{FieldMode2} for different time delays $\Delta$ with $\lambda=1.6\lambda_c$, $ E_{0}=0.5  E_{cr}$ , $\tau=10$m$^{-1}$.}
\label{DB-016}
\end{figure}

Now we turn to the following two same-signed Sauter pulses for the temporal profile of the electric field:
\begin{equation}\label{FieldMode2}
E\left(x,t\right) = E_{0} \exp \left(-\frac{x^{2}}{2 \lambda^{2}} \right ) \left( \text{sech}^{2}\left( \frac{t}{\tau} - \Delta \right)  + \text{sech}^{2}\left( \frac{t}{\tau} + \Delta \right ) \right) .
\end{equation}

Fig. \ref{Np-Double3Sauter} shows the (reduced) momentum spectrum for field form \eqref{FieldMode2}, where we can see the effects of spatial inhomogeneity with the decreasing value of $\lambda$ as merging of the two peaks present in the momentum spectrum, followed by the similar self-bunching pattern as in the single pulse case(Fig. \ref{self-bunching}), and the oscillatory pattern for $\lambda = 1.6 \lambda_c$.

The two peaks corresponding to the large values for $\lambda$ is expected, since it reflects the features of two same signed temporal Sauter pulses in the uniform field approximation.
However, we could notice that these two peaks are not symmetric as one would expect for the pure temporal case.
When the spatial extent of the electric field decreases, the particles created in the first Sauter pulse almost leaves the electric field and receives few accelerations from the second pulse, which creates the second bunch of particles with similar momentum as the first pulse thus leaving only one peak in the momentum spectrum.
With the further decrease in the value of $\lambda$, particles created from each Sauter pulses experience same self-bunching effects, while the two bunch of particles are separated spatially.
Comparing Fig. \ref{self-bunching} and Fig. \ref{Np-Double3Sauter}(b) for $\lambda = 10$ , $4$ and $3$, we can observe the same self-bunching distribution shape with the number density doubled for two Sauter pulses.

Interestingly, however, we observe oscillations in momentum spectrum for $\lambda \le 2 \lambda_c$, where for the same spatial extent we see no such effects in the single pulse case.
At such narrow focusing, in the semiclassical picture, one may expect that particles would leave the external field region with small momentum, and because of the temporal duration of the pulse, particles created continuously from the electric field would spread in the position space.
Since we have multiple pulses, particles from different sources are likely to have overlap in space and cause resonance in the momentum spectrum.
However, in this picture, we would only have small overlap in space for particles with vanishing momentum which contradicts with the oscillating curve in Fig. \ref{Np-Double3Sauter}(b).
It is possible that, at such extremely focused case, particles may be created via direct energy transfer.
And the resonance could then be explained by particles created from two pulses accumulating different phases in the process.

Moreover, the oscillation frequency in the momentum spectrum tends to increase with larger time delays $\Delta$ between two pulses, see Fig. \ref{DB-016}.
This fine structure in momentum spectrum for large $\Delta$ could be related to the time-energy uncertainty principle: large $\Delta$, being the longer characteristic time for energy change of the system, corresponds to more frequent changes in energy(momentum) of the system \cite{Kaminski:2018ywj}.

Normally, we do not expect interference-like patterns in symmetric pulses in spatially homogeneous scenarios where one can exclude the interference pattern using the turning point analysis, for more discussions see Refs. \cite{Akkermans:2011yn,Dumlu:2010ua,Dumlu:2011rr}.
However, it is shown in Ref. \cite{Kohlfurst:2012yw} that we could observe interference patterns for the symmetric pulse configurations if the pulses are short enough, and this is not apparent in the semiclassical pictures for the quantum nature of the process.
In the current case, we are observing strong nonlocal behavior of the produced particles \cite{Gies:2005bz,Lv:2018qxy} at extreme conditions that may cause resonance effects as a result of multiple pulses being sources for creation of particles which tend to manifest wavelike nature.

\subsection{Two anti-symmetric Sauter pulses}\label{result3}

\begin{figure}[h]\suppressfloats
\includegraphics[scale=0.6]{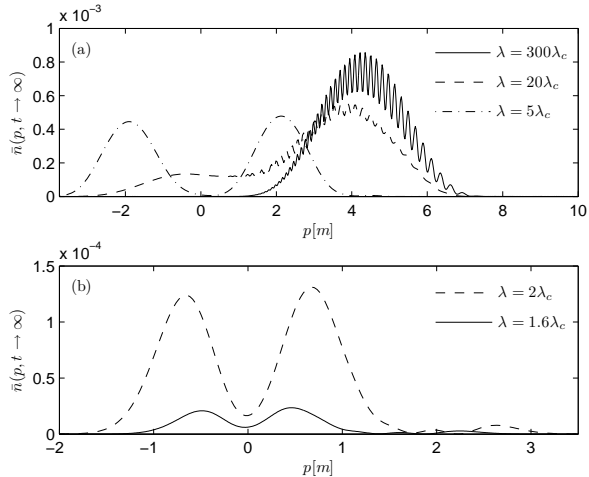}
\caption{Momentum spectrum for different spatial extents for two anti-symmetric pulses \eqref{FieldMode1} with $ E_{0}=0.5  E_{cr}$, $\tau=10$m$^{-1}$ and $\Delta=1$.}
\label{Np-Di1Sauter-new}
\end{figure}

At this point, it is worth turning our attention to temporal field forms that may induce interference effects by considering the following two anti-symmetric Sauter pulses:
\begin{equation}\label{FieldMode1}
E\left(x,t\right) =  E_{0} \exp \left(-\frac{x^{2}}{2 \lambda^{2}} \right ) \left( \text{sech}^{2}\left( \frac{t}{\tau} - \Delta \right)  - \text{sech}^{2}\left( \frac{t}{\tau} + \Delta \right ) \right) .
\end{equation}
The momentum spectrum for different values of $\lambda$ in \eqref{FieldMode1} is presented in Fig. \ref{Np-Di1Sauter-new}.
For $\lambda=300$ in Fig. \ref{Np-Di1Sauter-new} (a), we see the anticipated interference pattern as the signature of the anti-symmetric temporal pulses.
Furthermore, with the decrease of $\lambda$, we can see the interference effect is weaken and finally the momentum peak splits into two peaks with the shift towards the vanishing momentum.
This indicates that the finiteness of the spatial extent may weaken the signature of the temporal mode in the homogenous cases, and prevent the interference induced by the temporal structure(see also Ref. \cite{Aleksandrov:2017mtq}).
Notice that, the interference is not so pronounced even at the quasihomogeneous case which could be the consequence of introducing spatial variation into the external field as mentioned in Sec.~\ref{result1}.

The interference is related to the fact that one can not identify the source of the particles with the same momentum at large spatial extent\cite{Kohlfurst:2015zxi}.
However, when the spatial extent is small, some particles created from the first pulse will leave the electric field and can not be decelerated by the opposite signed pulse and continues to move on in the original direction ceasing interference with the second bunch.
Therefore, negative momentum peak appears corresponding to the first pulse.
With the further decrease of $\lambda$, the interference is terminated gradually and two peaks with opposite signs emerge in the momentum spectrum.
This is seen from Fig. \ref{Np-Di1Sauter-new} (b) for small values of $\lambda$, where the momentum spectrum is separate and is almost symmetric.

Same scenario is discussed in Ref. \cite{Kohlfurst:2015zxi}, yet we notice that at narrow focusing as shown in Fig. \ref{Np-Di1Sauter-new} (b), there may still be signs of oscillations around $p=2\sim3m$.
The quantum mechanical resonance effects which seem to occur for multiple temporal pulses with spatial focusing near the electron Compton wavelength is less pronounced for opposite signed pulses where the particles move in the opposite direction.

\subsection{Dynamically assisted pulse}\label{result4}

\begin{figure}[h]\suppressfloats
\includegraphics[scale=0.6]{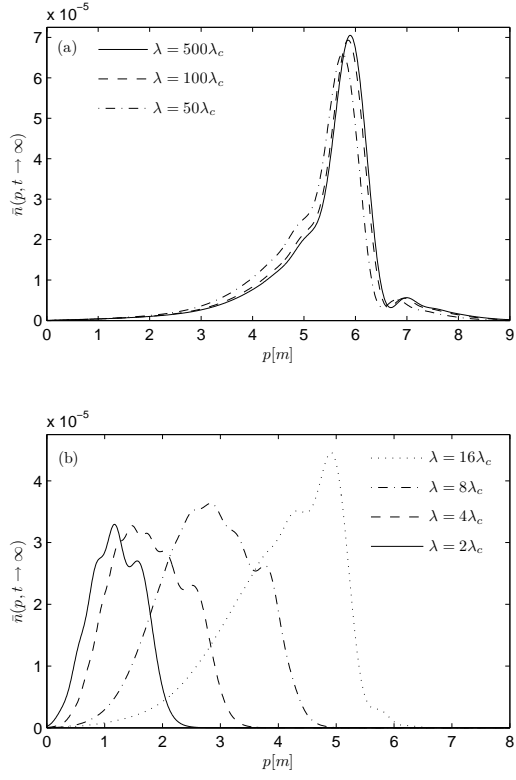}
\caption{Momentum spectrum for different spatial extents for the dynamically assisted pulse \eqref{FieldMode3} with $ E_{0}=0.3 E_{cr}$ , $\tau_1=20$m$^{-1}$, $\tau_2=1.4$m$^{-1}$ and $\epsilon=0.15$.}
\label{Np-DA}
\end{figure}

\begin{figure}[h]\suppressfloats
\includegraphics[scale=0.6]{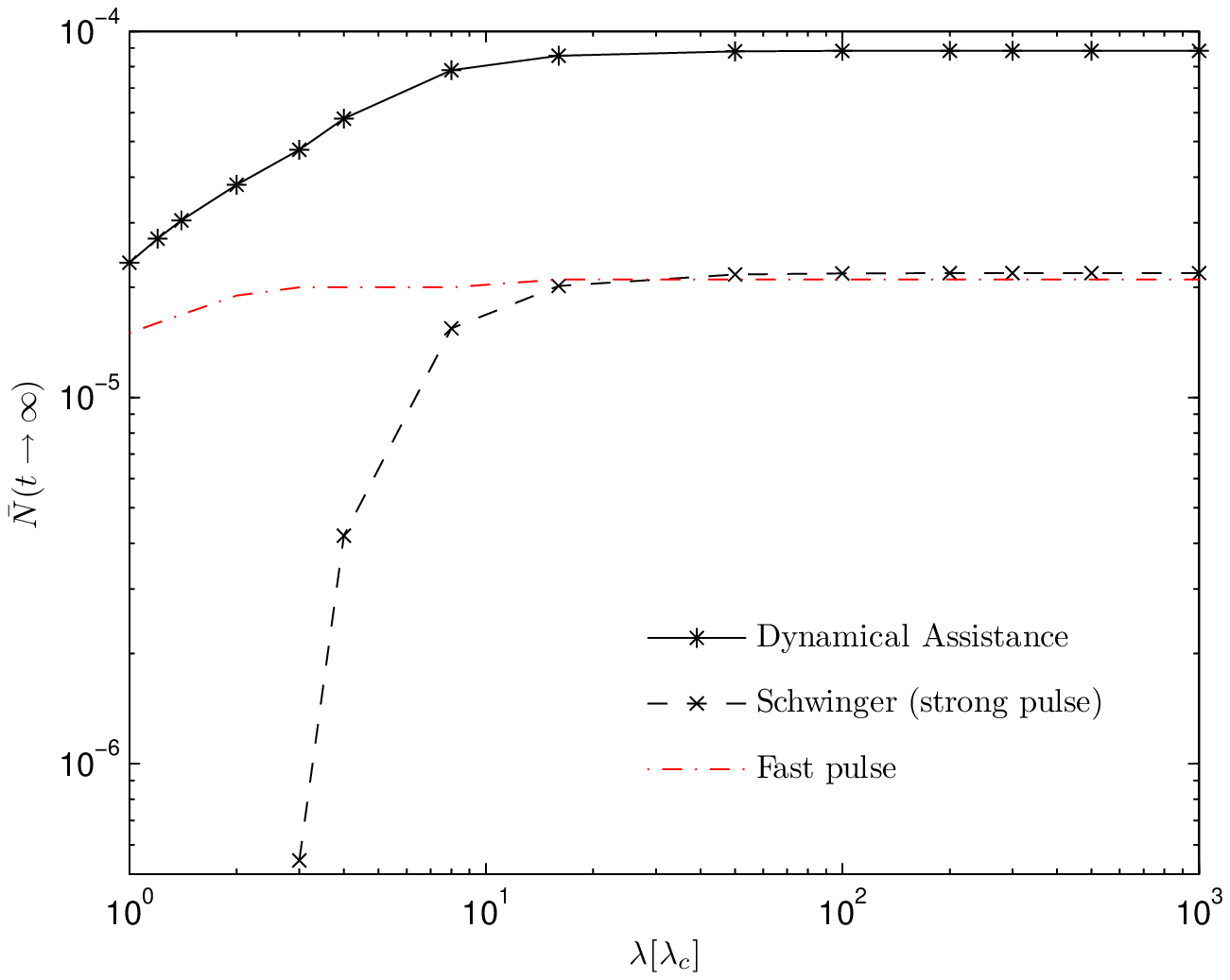}
\caption{Comparison of the reduced particle yield $\bar{N}(t\rightarrow\infty)$ for the dynamically assisted pulse, the strong pulse and the fast pulse at different spatial extents for the same field parameters as in Fig. \ref{Np-DA}. }
\label{DA-vs-non-DA}
\end{figure}

So far, we have investigated the effects of spatial inhomogeneity on interference signatures of various temporal fields.
It is also interesting to explore the effect of finite spatial extent of the electric field on the assistance mechanism using the following simple field form with numerically advantageous parameter choices:
\begin{equation}\label{FieldMode3}
E\left(x,t\right) = E_{0} \exp \left(-\frac{x^{2}}{2 \lambda^{2}} \right ) \left( \text{sech}^{2}\left( \frac{t}{\tau_1} \right)  + \epsilon \text{sech}^{2}\left( \frac{t}{\tau_2} \right ) \right) .
\end{equation}

In Fig. \ref{Np-DA}(a), the signature of the dynamical assisted pair creation is resumed for large values of $\lambda$.
However, in the momentum spectrum, we can not see the clear self-bunching pattern for decreasing values of $\lambda$, see Fig. \ref{Np-DA}(b).
Particularly, the momentum spectrum does not seem to vanish with the decrease of $\lambda$ as fast as it does for above mentioned scenarios.
It is because the contribution from the weak pulse is not weaken much by the strong spatial inhomogeneity.
Here, we have two temporal pulses with different time scales which are affected by the same spatial extent.
Therefore, when $\lambda$ is small, i.g., $\lambda \sim 10 \lambda_c$, it is small enough to reduce production rate concerning the strong pulse, but it still could be large regarding the fast pulse permitting assistance.

When $\lambda \ge 100 \lambda_c$, i.e., the quasihomogeneous regime, the momentum distribution is not symmetric compared to the pure temporal pulse results \cite{Orthaber:2011cm}.
This is again, caused by external field depending on space which takes the Gaussian shape as mentioned in Sec. \ref{result1} and Sec. \ref{result3}.
Since particles move in the field region, some of them may miss the exact timing for a kick by the fast pulse thus looses the symmetric pattern in the momentum spectrum.

These results are also captured if we calculate the total yield for different spatial extents, see Fig. \ref{DA-vs-non-DA}.
For $\lambda \ge 100 \lambda_c$, the reduced total yield does not change reflecting the uniform approximation for all combined or individual pulses.
Moreover, for small values of $\lambda$, the decrease of the total yield for the strong and fast pulses are understandable for the field energy is also decreasing with the small spatial extents.
However, the fast pulse plays a dominant role in the process and the dynamical assistance seems to be stronger such that the assistance is still present at the region where strong pulse is negligible.
We should also note that pair creation due to the fast pulse is not constrained by the minimum spatial extent given in Eq. \eqref{mini-lambda}, see red curve in Fig. \ref{DA-vs-non-DA}.
This is understandable, since the minimum estimate on $\lambda$ is obtained by requiring the work done by the external field over space is at least equal to two times the electron mass, which is valid for the tunneling process; however, in the case of the fast pulse, particles are created due to instant absorbtion of energy receiving relatively less constraint from the spatial focusing.

Spatial inhomogeneity is affecting two temporal pulses in different extents: the strong pulse experiences uniform approximation followed by the weakening caused by narrow spatial focusing; on the other hand, the fast pulse is hardly affected by spatial focusing which is still relatively large.
As a result, the combination of the two pulses--- the dynamical assistance, behaves as if we have stronger contributing fast pulse as $\lambda$ decrease.
The electric field parameters are chosen so that we display the effects of the overall spatial inhomogeneity on assistance mechanism with clear signatures in both momentum spectrum and total yield with less numerical constraints.
However, to perform complete study one needs to conduct further investigations by considering more realistic field shapes with larger parameter span.
Nonetheless, the calculations serve our purpose and the main argument in our article still holds.

\section{Conclusion}\label{summary}
In conclusion, we have studied the QED vacuum pair production in space and time dependent intense electric fields in 1+1 dimensions by
considering a simplified standing wave mode with different temporal structures while keeping the spatial part as Gaussian.
We have presented the effects of finite spatial extent in three regimes in terms of the spatial scale $\lambda$.
When spatial focusing is one order of magnitude greater than the temporal pulse size, we obtain the quasihomogeneous results where the reduced momentum spectrum and the total yield would no longer change with increasing $\lambda$, but the result may still be slightly different from the pure temporal case since some particles could always leave the electric field.
In the intermediate region where the focusing size is comparable to the temporal pulse, we observe weakening of the temporal pulse signatures: momentum spectrum merging for two same signed pulses; vanishing of the interference and peak splitting for opposite signed pulses.
For the extremely narrow spatial extent comparable to the Compton wavelength, we obtain the quantum resonance structure in momentum spectrum for two symmetric or antisymmetric pulses which is not present in the case of corresponding uniform electric fields or inhomogeneous fields with larger spatial extents.
It is interesting to relate such oscillatory behaviour to nonlocal feature of Schwinger pair production reported recently in computational quantum field theory approach to QED vacuum \cite{Lv:2018qxy}.
However, further explorations are needed to gain more insight on this resonance effect and provide quantitative explanation.

The inhomogeneity behaves differently in dynamical assisted Schwinger production where two different temporal scales are present.
While the total yield is reduced by the narrow spatial extent, the assistance is increasing significantly compared to the quasihomogeneous case for the spatial variation acts on the two temporal pulses with different extents.
Particularly, the extremely focused set up seems to make the nonlinear combination of dynamical particle creation and the Schwinger production most profound.

Theses results suggest that introducing spatial inhomogeneity is crucial when we consider realistic laser pulses.
The results arising from pure temporal considerations may be changed by the added spatial inhomogeneity, this is also emphasized in Ref. \cite{Aleksandrov:2017mtq}.
Detailed studies are needed for understanding the Schwinger pair production in inhomogeneous fields since the interplay between the time and space dependence of the external field tends to vary for different spatial scales with various temporal structures.
This further implies that more realistic laser pulses in 3+1 dimensions should be considered to obtain better understanding of vacuum pair production and provide more reliable predictions for future experiments.

\begin{acknowledgments}

\noindent
We would like to thank Obulkasim Olugh for useful discussion and valuable comments.
We are also grateful to Christian Kohlf\"{u}rst for his critical reading of the manuscript and important discussions.
This work was supported by the National Natural Science Foundation of China (NSFC) under
Grant No.\ 11875007, 11864039, 11564038 and 11475026.
The computation was carried out at the HSCC of the Beijing Normal University.

\end{acknowledgments}

\end{document}